\documentclass{article}
\def\Ref#1{(\ref{#1})}
\usepackage{amsmath}
\usepackage{amssymb}
\usepackage{cite}
\begin{document}
\begin{titlepage}
\noindent{\large \bf Exactly solvable models through the generalized empty
interval method, for multi-species interactions}

\vskip 2 cm

\begin{center}{A. Aghamohammadi$^{\rm a,d}${\footnote{mohamadi@azzahra.ac.ir}},
M.~Alimohammadi$^{\rm b}${\footnote{alimohmd@ut.ac.ir}}, \&
M.~Khorrami$^{\rm c,d}${\footnote {mamwad@iasbs.ac.ir}}
 } \vskip 5 mm

{\it{$^{\rm a}$ Department of Physics, Alzahra University,
             Tehran 19834, Iran. }

{   $^{\rm b}$ Physics Department, University of Tehran,
                North Karegar Avenue, Tehran, Iran. }

{ $^{\rm c}$ Institute for Advanced Studies in Basic Sciences,
             P.~O.~Box 159,\\ Zanjan 45195, Iran. }

{ $^{\rm d}$ Institute of Applied Physics, P. O. Box 5878
             Tehran 15875, Iran. }}

\end{center}

\begin{abstract}
\noindent Multi-species reaction-diffusion systems, with
nearest-neighbor interaction on a one-dimensional lattice are
considered. Necessary and sufficient constraints on the
interaction rates are obtained, that guarantee the closedness of
the time evolution equation for $E^{\mathbf a}_n(t)$'s, the
expectation value of the product of certain linear combination of
the number operators on $n$ consecutive sites at time $t$. The
constraints are solved for the single-species left-right-symmetric
systems. Also, examples of multi-species system for which the
evolution equations of $E^{\mathbf a}_n(t)$'s are closed, are
given.
\end{abstract}
\begin{center} {\bf PACS numbers:} 05.40.-a, 02.50.Ga

\end{center}

\end{titlepage}
\newpage
\section{Introduction}
There is a well-established framework for equilibrium statistical
mechanics, but thermal equilibrium is a special case, and there
isn't a corresponding straightforward framework for investigating
the properties of systems not in equilibrium. There is no general
approach to systems far from equilibrium. Different methods have
been used to study these models. These include analytical and
asymptotic methods, mean-field methods, and large-scale numerical
methods. For high-dimensional systems, mean-field techniques give
exact or reasonable approximate results. But their results for
low-dimensional systems are generally not adequate. So, people are
motivated to study stochastic models in low dimensions. Models in
low dimensions, should also be in principle easier to study. Exact
results for some models on a one-dimensional lattice have been
obtained, for example in
\cite{ScR,ADHR,KPWH,HS1,PCG,HOS1,HOS2,AL,AKK,RK,RK2,AKK2,AAMS,AM1,MA1,RK3,MAM}.

The term exactly-solvable have been used with different meanings.
In \cite{GS}, a ten-parameter family of reaction-diffusion
processes was introduced for which the evolution equation of
$n$-point functions contains only $n$- or less- point functions.
The average particle-number in each site has been obtained exactly
for these models. In \cite{AM2,MA2}, the same method has been used
to analyze the above mentioned ten-parameter family model on a
finite lattice with boundaries, and in \cite{SAK} similar method
has been used to study models with
next-nearest-neighbor-interactions. In \cite{AA} and \cite{RK3},
integrability means that the $N$-particle conditional
probabilities' S-matrix is factorized into a product of 2-particle
S-matrices. Another method which has been used to solve some
reaction diffusion models exactly is the empty interval method,
and its generalizations.

The empty interval method (EIM) has been used to analyze the one
dimensional dynamics of diffusion-limited coalescence
\cite{BDb,BDb1,BDb2,BDb3}. Using this method, the probability that
$n$ consecutive sites are empty has been calculated.  For the
cases of finite reaction-rates, some approximate solutions have
been obtained. EIM has been also generalized to study the kinetics
of the $q$-state one-dimensional Potts model in the
zero-temperature limit \cite{Mb}.

In \cite{AKA},  all the one dimensional reaction-diffusion models
with nearest neighbor interactions which can be exactly solved by
EIM have been studied. EIM has also been used to study a  model
with next nearest neighbor interaction \cite{HH}. In \cite{KAA},
exactly solvable models through the empty-interval method, for
more-than-two-site interactions were studied. In \cite{MB}, the
conventional EIM has been extended to a more generalized form.
Using this extended version, a model which can not solved by
conventional EIM has been studied.

In this article, we consider systems, in them particles of more
than one species interact on a one-dimensional lattice. The
interaction is nearest-neighbor. Each site of the lattice, either
is empty, or is occupies by one particle. In section 2, we seek
necessary and sufficient conditions on the reaction rates, so that
the time evolution equation for $E^{\mathbf a}_{k,n}(t)$ is
closed. This quantity is the expectation of the product of a
specific linear combination of the number operators (corresponding
to different species) at $n$ consecutive sites beginning from the
$k$-th site. In section 3, all single-species left-right symmetric
reaction-diffusion systems solvable through the generalized
empty-interval method, are classified. In section 4, multi-species
systems are investigated, which are solvable through the
generalized empty-interval method, but are effectively
single-species. In section 5, some specific families of
two-species systems are investigated, which are exactly solvable
through the generalized empty-interval method. Finally, section 6
is devoted to concluding remarks.
\section{Models solvable through the generalized empty-interval
method} To fix notations, let us briefly introduce the
multi-species reaction-diffusion systems with nearest-neighbor
interactions, on a periodic lattice. Let the lattice have $L+1$
sites. The observables of such a system are the operators
$N_i^\alpha$, where $i$ with $1\leq i\leq L+1$ denotes the site
number, and $\alpha$ with $1\leq\alpha\leq p+1$ denotes the type
of the particle. One can regard $\alpha=p+1$ as a vacancy.
$N^\alpha_i$ is equal to one, if the site $i$ is occupied by a
particle of type $\alpha$. Otherwise, $N^\alpha_i$ is zero. We
also have a constraint
\begin{equation}\label{1}
s_\alpha N^\alpha_i=1,
\end{equation}
where ${\mathbf s}$ is a covector the components of which
($s_\alpha$'s) are all equal to one. The constraint \Ref{1},
simply says that every site, either is occupied by a particle of
one type, or is empty. A representation for these observables is
\begin{equation}\label{2}
N_i^\alpha:=\underbrace{1\otimes\cdots\otimes 1}_{i-1}\otimes
N^\alpha\otimes\underbrace{1\otimes\cdots\otimes 1}_{L+1-i},
\end{equation}
where $N^\alpha$ is a diagonal $(p+1)\times(p+1)$ matrix the only
nonzero element of which is the $\alpha$'th diagonal element, and
the operators 1 in the above expression are also
$(p+1)\times(p+1)$ matrices. It is seen that the constraint
\Ref{1} can be written as
\begin{equation}\label{3}
{\mathbf s}\cdot{\mathbf N}=1,
\end{equation}
where ${\mathbf N}$ is a vector the components of which are
$N^\alpha$'s. The state of the system is characterized by a vector
\begin{equation}\label{4}
{\mathbf P}\in\underbrace{{\mathbb V}\otimes\cdots\otimes{\mathbb
V}}_{L+1},
\end{equation}
where ${\mathbb V}$ is a $(p+1)$-dimensional vector space. All the
elements of the vector ${\mathbf P}$ are nonnegative, and
\begin{equation}\label{5}
{\mathbf S}\cdot{\mathbf P}=1.
\end{equation}
Here ${\mathbf S}$ is the tensor-product of $L+1$ covectors
${\mathbf s}$.

As the number operators $N^\alpha_i$ are zero or one (and hence
idempotent), the most general observable of such a system is the
product of some of these number operators, or a sum of such terms.

The evolution of the state of the system is given by
\begin{equation}\label{6}
\dot{\mathbf P}={\mathcal H}\;{\mathbf P},
\end{equation}
where the Hamiltonian ${\mathcal H}$ is stochastic, by which it is
meant that its nondiagonal elements are nonnegative and
\begin{equation}\label{7}
{\mathbf S}\; {\mathcal H}=0.
\end{equation}
The interaction is nearest-neighbor, if the Hamiltonian is of the
form
\begin{equation}\label{8}
{\mathcal H}=\sum_{i=1}^{L+1}H_{i,i+1},
\end{equation}
where
\begin{equation}\label{9}
H_{i,i+1}:=\underbrace{1\otimes\cdots\otimes 1}_{i-1}\otimes H
\otimes\underbrace{1\otimes\cdots\otimes 1}_{L-i}.
\end{equation}
(It has been assumed that the sites of the system are identical,
that is, the system is translation-invariant. Otherwise $H$ in the
right-hand side of \Ref{9} would depend on $i$.) The two-site
Hamiltonian $H$ is stochastic, that is, its non-diagonal elements
are nonnegative, and the sum of the elements of each of its
columns vanishes:
\begin{equation}\label{10}
({\mathbf s}\otimes{\mathbf s})H=0.
\end{equation}

Now consider a certain class of such observables, namely
\begin{equation}\label{11}
{\mathcal E}^{\mathbf a}_{k,n}:=\prod_{l=k}^{k+n-1}({\mathbf
a}\cdot{\mathbf N}_l),
\end{equation}
where ${\mathbf a}$ is a specific $(p+1)$-dimensional covector,
and ${\mathbf N}_i$ is a vector the components of which are the
operators $N_i^\alpha$. We want to find criteria for $H$, so that
the evolutions of the expectations of ${\mathcal E}^{\mathbf
a}_{k,n}$'s are closed, that is the time-derivative of their
expectation is expressible in terms of the expectations of
${\mathcal E}^{\mathbf a}_{k,n}$'s themselves. Denoting the
expectations of these observables by $E^{\mathbf a}_{k,n}$,
\begin{equation}\label{12}
E^{\mathbf a}_{k,n}:={\mathbf S}\;{\mathcal E}^{\mathbf
a}_{k,n}{\mathbf P},
\end{equation}
we have
\begin{align}\label{13}
\dot E^{\mathbf a}_{k,n}=&{\mathbf S}\;{\mathcal E}^{\mathbf
a}_{k,n}
{\mathcal H}\;{\mathbf P},\nonumber\\
=&{\mathbf S}\;{\mathcal E}^{\mathbf a}_{k,n}H_{k-1,k}\;{\mathbf P}\nonumber\\
&+\sum_{l=1}^{n-1}{\mathbf S}\;{\mathcal E}^{\mathbf
a}_{k,n}H_{k-1+l,k+l}
\;{\mathbf P}\nonumber\\
&+{\mathbf S}\;{\mathcal E}^{\mathbf
a}_{k,n}H_{k+n-1,k+n}\;{\mathbf P}.
\end{align}
From this, and using \Ref{3}, it is seen that the necessary and
sufficient conditions that the left-hand side be expressible in
terms of $E^{\mathbf a}_{k,n}$'s are
\begin{align}\label{14-16}
({\mathbf s}\otimes{\mathbf s})[({\mathbf s}\cdot{\mathbf N})
\otimes({\mathbf a}\cdot{\mathbf N})]H =&\mu_{\mathrm L} ({\mathbf
s}\otimes{\mathbf s})[({\mathbf s}\cdot{\mathbf N})
\otimes({\mathbf a}\cdot{\mathbf N})]\nonumber\\
&+\theta_{\mathrm L} ({\mathbf s}\otimes{\mathbf s})[({\mathbf
s}\cdot{\mathbf N}) \otimes({\mathbf s}\cdot{\mathbf N})]
\nonumber\\
&+\nu_{\mathrm L} ({\mathbf s}\otimes{\mathbf s})[({\mathbf
a}\cdot{\mathbf N}) \otimes({\mathbf a}\cdot{\mathbf N})],
\\
({\mathbf s}\otimes{\mathbf s})[({\mathbf a}\cdot{\mathbf N})
\otimes({\mathbf a}\cdot{\mathbf N})]H =&\lambda ({\mathbf
s}\otimes{\mathbf s})[({\mathbf a}\cdot{\mathbf N})
\otimes({\mathbf a}\cdot{\mathbf N})],\\
({\mathbf s}\otimes{\mathbf s})[({\mathbf a}\cdot{\mathbf N})
\otimes({\mathbf s}\cdot{\mathbf N})]H =&\mu_{\mathrm R} ({\mathbf
s}\otimes{\mathbf s})[({\mathbf a}\cdot{\mathbf N})
\otimes({\mathbf s}\cdot{\mathbf N})]\nonumber\\
&+\theta_{\mathrm R} ({\mathbf s}\otimes{\mathbf s})[({\mathbf
s}\cdot{\mathbf N}) \otimes({\mathbf s}\cdot{\mathbf N})]
\nonumber\\
&+\nu_{\mathrm R} ({\mathbf s}\otimes{\mathbf s})[({\mathbf
a}\cdot{\mathbf N}) \otimes({\mathbf a}\cdot{\mathbf N})],
\end{align}
for some arbitrary numbers $\lambda$, $\mu_{\mathrm L,R}$,
$\theta_{\mathrm L,R}$, and $\nu_{\mathrm L,R}$. Using the
identity
\begin{equation}\label{17}
{\mathbf s}({\mathbf b}\cdot{\mathbf N})={\mathbf b},
\end{equation}
for an arbitrary covector ${\mathbf b}$, one arrives at
\begin{align}\label{18}
({\mathbf s}\otimes{\mathbf a})H=&\mu_{\mathrm L} ({\mathbf
s}\otimes{\mathbf a})+\theta_{\mathrm L} ({\mathbf
s}\otimes{\mathbf s})+\nu_{\mathrm L}
({\mathbf a}\otimes{\mathbf a}),\nonumber\\
({\mathbf a}\otimes{\mathbf a})H=&\lambda ({\mathbf
a}\otimes{\mathbf a}),\nonumber\\
({\mathbf a}\otimes{\mathbf s})H=&\mu_{\mathrm R} ({\mathbf
a}\otimes{\mathbf s})+\theta_{\mathrm R} ({\mathbf
s}\otimes{\mathbf s})+\nu_{\mathrm R} ({\mathbf a}\otimes{\mathbf
a}).
\end{align}
The Hamiltonian $H$ should of course satisfy \Ref{10} as well, and
its nondiagonal elements should be nonnegative. In this case, the
evolution equation for $E^{\mathbf a}_{k,n}$, for $0<n<L+1$,
becomes
\begin{equation}\label{19}
\dot E^{\mathbf a}_{k,n}=\theta_{\mathrm L}E^{\mathbf a}_{k+1,n-1}
+\nu_{\mathrm L}E^{\mathbf a}_{k-1,n+1} +\theta_{\mathrm
R}E^{\mathbf a}_{k,n-1} +\nu_{\mathrm R}E^{\mathbf a}_{k,n+1}
+[\mu_{\mathrm L}+(n-1)\lambda+\mu_{\mathrm R}]E^{\mathbf
a}_{k,n}.
\end{equation}
For $n=L+1$, one has
\begin{equation}\label{20}
\dot E^{\mathbf a}_{1,L+1}=(L+1)\lambda E^{\mathbf a}_{1,L+1}.
\end{equation}
For $n=0$, from \Ref{5} we have the boundary condition
\begin{equation}\label{21}
E^{\mathbf a}_{k,0}=1.
\end{equation}
Equations \Ref{19} and \Ref{20}, and the boundary condition
\Ref{21} are a closed set of evolution equations for $E^{\mathbf
a}_{i,j}$'s. These equations are quite similar to those obtained
in \cite{AKA}. In fact, the case there is a special case of what
considered here, with $p=1$ and ${\mathbf a}=(0,1)$. Although the
criterion for the closedness of the evolution equations does
depend on $p$ and ${\mathbf a}$, the evolution equations for
$E^{\mathbf a}_{i,j}$'s do not depend on these. In \cite{AKA},
situations were considered in which $\lambda$ was zero, so that
the evolution of the block comes solely from its ends. This makes
solving the evolution equations easier. Finally, since the system
under consideration is translationally-symmetric, if the initial
condition is translationally-invariant, the state of the system
would remain translationally-invariant at all times. In this case,
$E^{\mathbf a}_{k,n}$ does not depend on $k$, and one would have
\begin{align}\label{22}
\dot E^{\mathbf a}_n=&(\theta_{\mathrm L}+\theta_{\mathrm R})
E^{\mathbf a}_{n-1} +(\nu_{\mathrm L}+\nu_{\mathrm R})E^{\mathbf
a}_{n+1} +[\mu_{\mathrm L}+(n-1)\lambda+\mu_{\mathrm R})E^{\mathbf
a}_n,\quad 0<n<L+1,\nonumber\\
\dot E^{\mathbf a}_{L+1}=&(L+1)\lambda E^{\mathbf
a}_{L+1},\nonumber\\
E^{\mathbf a}_0=&1.
\end{align}

The system is left-right symmetric, iff the Hamiltonian is
invariant under permutation:
\begin{equation}\label{23}
\Pi\; H\;\Pi=H,
\end{equation}
where $\Pi$ is the permutation matrix:
\begin{equation}\label{24}
\Pi({\mathbf u}\otimes{\mathbf v})={\mathbf v}\otimes{\mathbf u}.
\end{equation}
It is easily seen that if $H$ satisfies \Ref{23} and \Ref{18},
then
\begin{equation}\label{25}
\mu_{\mathrm L}=\mu_{\mathrm R}=\mu,\quad \nu_{\mathrm
L}=\nu_{\mathrm R}=\nu,\quad \theta_{\mathrm L}=\theta_{\mathrm
R}=\theta.
\end{equation}
\section{Classification of the single-species left-right-symmetric
reaction-diffusion systems, which are solvable through the
generalized empty-interval method} For a single-species system,
the vector space ${\mathbb V}$ is two-dimensional. Take a covector
${\mathbf a}$, which is not a multiple of ${\mathbf s}$. (The case
${\mathbf a}$ proportional to ${\mathbf s}$ is trivial, as
${\mathbf s}\cdot{\mathbf N}=1$). The set $B:=\{{\mathbf
a}\otimes{\mathbf a},{\mathbf a}\otimes{\mathbf s},{\mathbf
s}\otimes{\mathbf a},{\mathbf s}\otimes{\mathbf s}\}$, is a basis
for ${\mathbb V}\otimes{\mathbb V}$. If $H$ satisfies \Ref{18}
(and of course \Ref{10}), then one has the matrix elements of the
Hamiltonian in this basis. However, not every Hamiltonian in this
form represent a stochastic system. The nondiagonal elements of
the Hamiltonian in the physical (ordinary) basis should be
nonnegative. So, for the single-species systems, the task of
classifying the systems solvable through the generalized
empty-interval method, reduces to finding the matrix elements of
the Hamiltonian in the physical basis (in terms of the covector
${\mathbf a}$, and the scalars $\lambda$, $\mu$, $\nu$, and
$\theta$); and imposing the criterion that the nondiagonal
elements of $H$ in the physical basis be nonnegative. Taking the
covector ${\mathbf a}$ like
\begin{equation}\label{26}
{\mathbf a}=(a_1,a_2),
\end{equation}
and noting that $a_1\ne a_2$ (otherwise ${\mathbf a}$ would be
proportional to ${\mathbf s}$), and that one can rescale ${\mathbf
a}$ without changing the conditions \Ref{18}, it is seen that one
can take
\begin{equation}\label{27}
a_1=\xi+1,\quad a_2=\xi-1,
\end{equation}
without loss of generality. Then, imposing the criterion that the
nondiagonal elements of the Hamiltonian are nonnegative, and
assuming left-right symmetry, one arrives at the following set of
inequalities.
\begin{align}\label{28}
(1+\xi^2)\Lambda+2\xi^2\mu+2\xi\theta&\geq|2\xi(\Lambda+\mu)|,\nonumber\\
-(1+\xi^2)(2\mu +\Lambda)-4\xi\nu-2\xi\theta&\geq
|2[(1+\xi^2)\nu+\xi(\Lambda+2\mu)+\theta]|,\nonumber\\
-2\mu\geq (1-\xi^2)\Lambda-2\xi^2\mu-2\xi\theta&\geq
|2[(\xi^2-1)\nu+\xi\mu+\theta]|,
\end{align}
where
\begin{equation}\label{29}
\Lambda:=-\lambda+2\xi\nu.
\end{equation}
We also define
\begin{align}\label{32}
\tau:=&\theta-\xi,\nonumber\\
\Lambda^{\pm}:=&\Lambda\pm 2\nu.
\end{align}
We have to solve the inequalities \Ref{28}, for a given value of
$\xi$. It is seen that changing the signs of $\xi$, $\nu$, and
$\theta$ simultaneously, while keeping the signs of $\lambda$,
$\mu$, and $\Lambda$ fixed, does not change the inequalities. So
it is sufficient to solve the inequalities for nonnegative $\xi$.

The detailed calculations can be found in the appendix. The results
are the following.\\
{\textbf{i)}}
\begin{equation}\label{2030}
\xi=1,\quad\mu=\theta=0,\quad\nu\leq 0, \quad 4\nu\leq\lambda\leq
2\nu.
\end{equation}
The reactions for systems in this class are
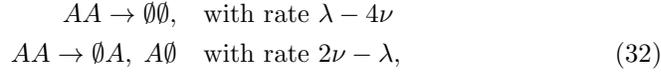
\begin{align}\label{II1}
  AA\to\emptyset\emptyset,\quad &\hbox{with rate }\lambda-4\nu\nonumber\\
  AA\to\emptyset A,\; A\emptyset\quad &\hbox{with rate }2\nu-\lambda,
\end{align}
for which the evolution equation of $\langle n_1\cdots n_k\rangle$
is closed, where $n_i$ is the number operator at the site $i$.\\
\\
{\textbf{ii)}}
\begin{equation}\label{2035}
0<\xi<1,\quad\mu<0\;(\mu=-1)\quad(\Lambda^-,\Lambda^+)\hbox{
inside the tetragon }ABCE,\quad\tau\hbox{ satisfies \Ref{33}}.
\end{equation}
The coordinates of the vertices of the tetragon $ABCE$ are
$$A\Big(-{{1-\xi}\over{1+\xi}},1\Big),\quad B\Big(1,
{{3+\xi}\over{1+\xi}}\Big),\quad
C\Big({{3-\xi}\over{1-\xi}},1\Big),\quad
E\Big({{1+6\xi-3\xi^2}\over{(1-\xi)(1+3\xi)}},
-{{1+3\xi^2}\over{(1-\xi)(1+3\xi)}}\Big),$$
where the first coordinate
is $\Lambda^-$, and the second is $\Lambda^+$. As an example in this class, take
$\xi=1/2$, and $\Lambda^+=\Lambda^-=1$, which lead to a system with
following interactions.
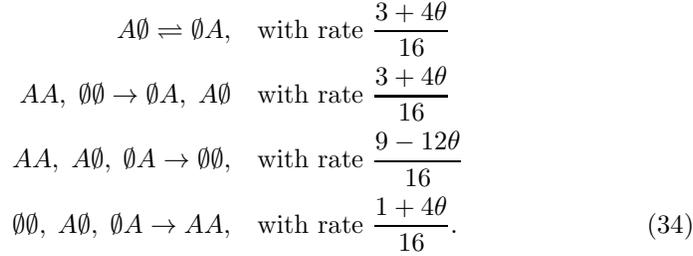
\begin{align}\label{II2}
  A\emptyset\rightleftharpoons\emptyset A,\quad &\hbox{with rate }\frac{3+4\theta}{16}\nonumber\\
  AA,\;\emptyset\emptyset\to\emptyset A,\;A\emptyset\quad &\hbox{with rate }
  \frac{3+4\theta}{16}\nonumber\\
  AA,\;A\emptyset,\;\emptyset A\to\emptyset\emptyset,\quad &\hbox{with rate }
  \frac{9-12\theta}{16}\nonumber\\
  \emptyset\emptyset,\;A\emptyset,\;\emptyset A\to AA,\quad &\hbox{with rate }
  \frac{1+4\theta}{16}.
\end{align}
For this system, the evolution equation of
$\langle[2n_1-(1/2)]\cdots[2n_k-(1/2)]\rangle$ is closed.\\
\\
{\textbf{iii)}}
\begin{equation}\label{2042}
1<\xi,\quad\mu<0\;(\mu=-1),\quad \tau=-{{\xi^2+1}\over{2\xi}},\quad
\Lambda^{\pm}={{\xi\pm 1}\over\xi}.
\end{equation}
It is seen that here there is no allowed region for the rates, but
a single point (apart from scaling) for each value of $\xi$. Systems
in this class, correspond to the reactions
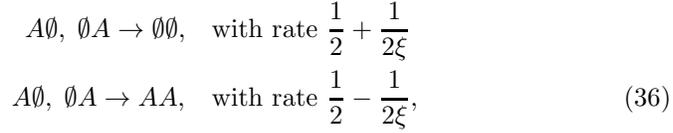
\begin{align}\label{II3}
  A\emptyset,\;\emptyset A\to\emptyset\emptyset,\quad &\hbox{with rate }
  \frac{1}{2}+\frac{1}{2\xi}\nonumber\\
  A\emptyset,\;\emptyset A\to AA,\quad &\hbox{with rate }
  \frac{1}{2}-\frac{1}{2\xi},
\end{align}
and for them the evolution equation of
$\langle(2n_1+\xi-1)\cdots(2n_k+\xi-1)\rangle$ is closed.\\
\\
{\textbf{iv)}}
\begin{align}\label{2044}
&\xi=0,\quad \mu\ne 0\;\;(\mu=-1),\quad
(\Lambda^-,\Lambda^+)\hbox{ is inside the square
}A_0B_0C_0D_0.\nonumber\\
&\tau\hbox{ satisfies \Ref{31}.}
\end{align}
The vertices of the tetragon $A_0B_0C_0D_0$ are
$$A_0(-1,1),\quad B_0(1,3),\quad C_0(3,1),\quad D_0(1,-1).$$\\
As an example in this class, take $\Lambda^+=\Lambda^-=1$, which
lead to a system with following interactions.
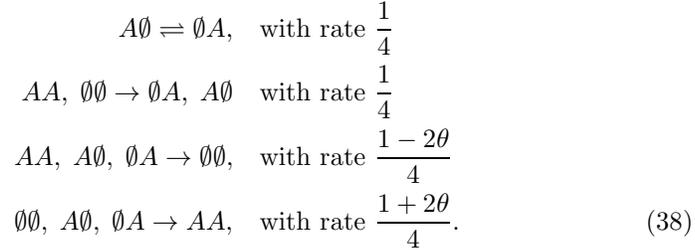
\begin{align}\label{II4}
  A\emptyset\rightleftharpoons\emptyset A,\quad &\hbox{with rate }
  \frac{1}{4}\nonumber\\
  AA,\;\emptyset\emptyset\to\emptyset A,\;A\emptyset\quad &\hbox{with rate }
  \frac{1}{4}\nonumber\\
  AA,\;A\emptyset,\;\emptyset A\to\emptyset\emptyset,\quad &\hbox{with rate }
  \frac{1-2\theta}{4}\nonumber\\
  \emptyset\emptyset,\;A\emptyset,\;\emptyset A\to AA,\quad &\hbox{with rate }
  \frac{1+2\theta}{4}.
\end{align}
For this system, the evolution equation of
$\langle(2n_1-1)\cdots\otimes(2n_k-1)\rangle$ is closed.\\
\\
{\textbf{v)}}
\begin{equation}\label{2046}
\xi=1,\quad \mu\ne 0\;\;(\mu=-1),\quad (\Lambda^-,\Lambda^+)\hbox{
is inside }{\mathbb S}.\quad \tau\hbox{ satisfies \Ref{45}.}
\end{equation}
The region $\mathbb{S}$, is the region limited by the lines
$A_1B'$, the horizontal line passing through $B'$, and the
line passing through $A_1$ with the slope $-1$, containing
the point $(1,1)$, where $$A_1(0,1),\quad B'(0,2).$$\\
As an example in this class, take $\Lambda^+=\Lambda^-=1$, which
lead to a system with following interactions.
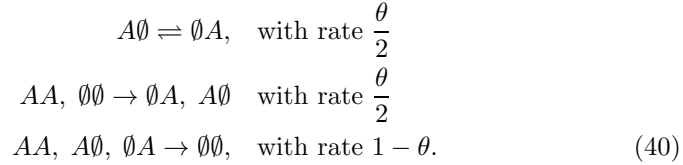
\begin{align}\label{II5}
  A\emptyset\rightleftharpoons\emptyset A,\quad &\hbox{with rate }
  \frac{\theta}{2}\nonumber\\
  AA,\;\emptyset\emptyset\to\emptyset A,\;A\emptyset\quad &\hbox{with rate }
  \frac{\theta}{2}\nonumber\\
  AA,\;A\emptyset,\;\emptyset A\to\emptyset\emptyset,\quad &\hbox{with rate }
  1-\theta.
\end{align}
For this system, the evolution equation of $\langle n_1\cdots
n_k\rangle$ is closed.
\section{Effectively single-species systems}
For a specific $(p+1)$-dimensional covector, we seek Hamiltonians
satisfying \Ref{18} with some values of $\lambda$, $\mu_{\mathrm
L,R}$, $\nu_{\mathrm L,R}$, and $\theta_{\mathrm L,R}$. There are
cases, however, where $p$-species systems are effectively
single-species. Suppose we can decompose the states of the system
into two subsets 1 and 2. Corresponding to these states, one
defines the covectors ${\mathbf E}^1$, and ${\mathbf E}^2$.
${\mathbf E}^1$, for example, is the sum of the covectors
corresponding to the states belonging to the first subset (the
microstates of the first state). It is clear that
\begin{equation}\label{47}
{\mathbf E}^1+{\mathbf E}^2={\mathbf s}.
\end{equation}
To have an effectively two-state system, the probability that the
system goes from one of the microstates of the state $i$ to the
state $j$, should not depend on the microstates. In terms of the
Hamiltonian, this means
\begin{equation}\label{48}
{\mathbf E}^i\; H=\sum_j\tilde H^i{}_j {\mathbf E}^j.
\end{equation}
In this case, the system described by the Hamiltonian $\tilde H$
is a two-state system. If, moreover, one seeks informations about
the original system, which are expressible in terms of only the
states 1 and 2 (and not dependent on the microstates) then the
system is an effectively two-state system. For example, if the
states of a system are white, blue, and red, one can define the
states white and colored, provided the probability that the system
changes from red to white is the same as that of changing from
blue to white. If in addition, we are only interested in
probabilities of the system being white or colored, then the
system is effectively a two-state system. Obvious generalizations
of these systems with nearest-neighbor interactions on a lattice,
are systems for which
\begin{equation}\label{49}
({\mathbf E}^i\otimes{\mathbf E}^j)\; H=\sum_{k,l} \tilde
H^{ij}{}_{kl}({\mathbf E}^k\otimes{\mathbf E}^l).
\end{equation}

An example is a system consisting of two kinds of particles ($A$
and $B$) diffusing on a lattice:
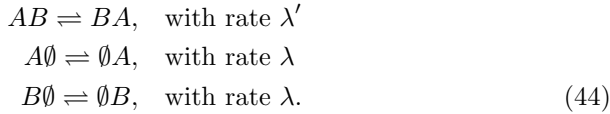
\begin{align}\label{2053}
  AB\rightleftharpoons BA,\quad &\hbox{with rate }\lambda'\nonumber\\
  A\emptyset\rightleftharpoons \emptyset A,\quad &\hbox{with rate }\lambda\nonumber\\
  B\emptyset\rightleftharpoons \emptyset B,\quad &\hbox{with rate }\lambda.
\end{align}
This system is effectively one-species, as far as one is concerned only with
probabilities of finding particles (not of a specific type).

Now let us return to the problem of finding solutions to \Ref{18},
with a specified covector ${\mathbf a}$. We want to show that if
the components of the covector ${\mathbf a}$ take only two values,
then the system under consideration is an effectively two-state
system (or an effectively single-species system, with the states
occupied and empty). If the components of ${\mathbf a}$ take only
two values, then one can write ${\mathbf a}$ as
\begin{equation}\label{50}
{\mathbf a}=a_1{\mathbf E}^1+a_2{\mathbf E}^2,
\end{equation}
where ${\mathbf E}^i$'s are covectors with the property that their
components are either zero or one, and their sum is equal to
${\mathbf s}$. It is seen that the covectors ${\mathbf s}$ and
${\mathbf a}$ are linear combinations of ${\mathbf E}^1$ and
${\mathbf E}^2$, and vice versa. So a result of equations \Ref{18}
and \Ref{10} is that $({\mathbf E}^i\otimes{\mathbf E}^j)H$ is a
linear combination of $({\mathbf E}^k\otimes{\mathbf E}^l)$'s;
that is, \Ref{49} holds. Also, one notes that
\begin{equation}\label{51}
{\mathbf a}\cdot{\mathbf N}=a_1{\mathbf E}^1\cdot{\mathbf
N}+a_2{\mathbf E}^2\cdot{\mathbf N},
\end{equation}
which means that the information we seek involves the
probabilities corresponding to only the subsets 1 and 2. So,
multi-species systems solvable through the generalized
empty-interval method, with covectors the components of which take
only two values, are effectively single-species.
\section{Some two-species examples}
The states of a two-species system on each site of the lattice can
be represented by $A$, $B$, and $\emptyset$, the latter being a
vacancy. As the first example, consider a system for which the
Hamiltonian is symmetric, which means the rate of each reaction is
equal to the rate of its reverse reaction. Also let $\lambda=0$ in
\Ref{18}. For the three-dimensional covector ${\mathbf a}$, take
the choice
\begin{equation}\label{52}
{\mathbf a}=(1,-1,0).
\end{equation}
One can then solve \Ref{18} and \Ref{10} for $H$ and $\mu_{\mathrm
L,R}$, $\nu_{\mathrm L,R}$, and $\theta_{\mathrm L,R}$. Also the
rates (the nondiagonal elements of $H$) should be nonnegative.
These constraints lead to a system with the following reaction
rates.
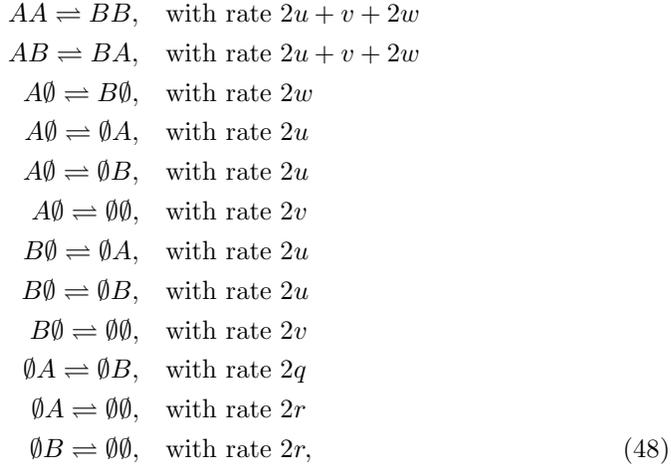
\begin{align}\label{53}
  AA\rightleftharpoons BB,\quad &\hbox{with rate }2u+v+2w\nonumber\\
  AB\rightleftharpoons BA,\quad &\hbox{with rate }2u+v+2w\nonumber\\
  A\emptyset\rightleftharpoons B\emptyset,\quad &\hbox{with rate }2w\nonumber\\
  A\emptyset\rightleftharpoons \emptyset A,\quad &\hbox{with rate }2u\nonumber\\
  A\emptyset\rightleftharpoons \emptyset B,\quad &\hbox{with rate }2u\nonumber\\
  A\emptyset\rightleftharpoons \emptyset\emptyset,\quad &\hbox{with rate }2v\nonumber\\
  B\emptyset\rightleftharpoons \emptyset A,\quad &\hbox{with rate }2u\nonumber\\
  B\emptyset\rightleftharpoons \emptyset B,\quad &\hbox{with rate }2u\nonumber\\
  B\emptyset\rightleftharpoons \emptyset\emptyset,\quad &\hbox{with rate }2v\nonumber\\
  \emptyset A\rightleftharpoons \emptyset B,\quad &\hbox{with rate }2q\nonumber\\
  \emptyset A\rightleftharpoons \emptyset\emptyset,\quad &\hbox{with rate }2r\nonumber\\
  \emptyset B\rightleftharpoons \emptyset\emptyset,\quad &\hbox{with rate
  }2r,
\end{align}
with the condition
\begin{equation}\label{54}
r+2q=v+2w.
\end{equation}
For these rates,
\begin{align}\label{55}
\mu_{\mathrm L}=\mu_{\mathrm R}&=-(4u+2v+4w),\nonumber\\
\nu_{\mathrm L}=\nu_{\mathrm R}=\theta_{\mathrm L}=\theta_{\mathrm
R}&=0.
\end{align}
In this example, the evolution of $E^{\mathbf a}_n$'s are very
simple. In fact the evolution equations decouple and one has
\begin{align}\label{56}
\dot E^{\mathbf a}_n&=2\mu E^{\mathbf a}_n,\quad
0<n<L+1,\nonumber\\
\dot E^{\mathbf a}_{L+1}&=0,\nonumber\\
E^{\mathbf a}_0&=1.
\end{align}

The second example is less trivial. Let $H$ satisfy \Ref{23}, that
is the system has left-right symmetry. Also let the covector
${\mathbf a}$ be
\begin{equation}\label{57}
{\mathbf a}=(1,\xi,0),
\end{equation}
and $\lambda=0$. From the first (or third) equations of \Ref{18},
one can find $\mu$, $\nu$, and $\theta$ in terms of the rates and
the parameter $\xi$. The remaining equations are linear equations
for rates. However, we have inequalities as well, namely the rates
should be nonnegative. A specific class of the solutions can be
obtained as
\begin{equation}\label{58}
H=\begin{pmatrix}
           0&b&0&b&0&0&0&0&0\\
           0&D_2&0&0&0&0&0&0&0\\
           0&c&D_3&0&0&r&\xi g&\xi(g+q)&\xi(g+q)\\
           0&0&0&D_4&0&0&0&0&0\\
           0&d&0&d&0&0&0&0&0\\
           0&0&g&f&0&D_6&q&0&0\\
           0&0&\xi g&c&0&\xi(g+q)&D_7&r&\xi(g+q)\\
           0&f&q&0&0&0&g&D_8&0\\
           0&0&0&0&0&u&0&u&D_9
           \end{pmatrix},
\end{equation}
where each diagonal element is minus the sum of the other elements
in its column, and the following relations hold between the rates.
\begin{align}\label{59}
(1-\xi)b&=\xi[c+f+(1-\xi)d],\nonumber\\
(1-\xi)(g+q)&=f+(1-\xi)d,\nonumber\\
\xi u&=(1-\xi)r.
\end{align}
It is clear that for $0<\xi<1$, one can always find rates
satisfying \Ref{59}. With these rates, one has
\begin{align}\label{60}
\mu&=-(1+\xi)(g+q),\nonumber\\
\nu&=g+q,\nonumber\\
\theta&=\xi(g+q).
\end{align}
\section{Concluding remarks}
Among the aims of investigating reaction-diffusion systems is to
find as many $n$-point functions (of number operators) as
possible. There are systems for which one can find certain classes
of these correlators.

The empty interval method was first introduced to investigate the
probability of finding $n$ neighboring empty sites, for systems
consisting of one species, with nearest-neighbor interactions. One
can generalize this method, in several aspects. One way is to
consider systems with more-than-two-site interactions. Another way
is to consider multi-species systems, and ask for the expectation
of the product of certain linear combinations of the number
operators. While this does not (generally) give the densities of
each specific kind of particles, it does give a certain
combination of these densities.

What was introduced here, was a set of constraints the
reaction-diffusion systems should satisfy, in order that the
system be solvable through the generalization of the empty
interval method. For example, in a system consisting of particles
of one kind (with nearest-neighbor interaction), there are 12
independent reaction-rates. If one demands that the system be
left-right symmetric, the number of independent rates is reduced
to 7. Among these, there exists a 5-parameter family, the models
of which are solvable through the generalized empty interval
method. The classification of this family, and some examples, were
discussed in section 3. An interesting problem may be to classify
the models satisfying the solvability conditions for multi-species
systems.

\section{Appendix}
We want to solve the inequalities \Ref{28}, for a given value of
$\xi$. It is seen that changing the signs of $\xi$, $\nu$, and
$\theta$ simultaneously, while keeping the signs of $\lambda$,
$\mu$, and $\Lambda$ fixed, does not change the inequalities. So
it is sufficient to solve the inequalities for nonnegative $\xi$.

First consider the case $\mu=0$. If $\mu=0$, and $\xi=1$, then one
arrives at\\
{\textbf{i)}}
\begin{equation}\label{30}
\xi=1,\quad\mu=\theta=0,\quad\nu\leq 0, \quad 4\nu\leq\lambda\leq
2\nu.
\end{equation}

Another case is $\mu=0$, $\xi\ne 1$. One can consider the subcases
$\xi=0$ and $\xi\ne 0$, and show that in both subcases,
$\Lambda=\theta=\nu=0$, which means the Hamiltonian is zero. So,
if $\xi\ne 1$, for any nontrivial solution, $\mu\ne 0$.

If $\mu\ne 0$, then it should be negative. One can divide the
Hamiltonian by $-\mu$ (which is like taking $\mu=-1$). The
inequalities \Ref{18} are then rewritten as
\begin{align}\label{31}
(1-\xi)[-(1-\xi)\Lambda^- +2(1+\tau)]&\geq 0,\nonumber\\
(1-\xi)[(1+\xi)\Lambda^- +2\tau]&\geq 0,\nonumber\\
(1+\xi)[-(1+\xi)\Lambda^+ +2(1-\tau)]&\geq 0,\nonumber\\
(1+\xi)[(1-\xi)\Lambda^- -2\tau]&\geq 0,\nonumber\\
(1+\xi)^2\Lambda-2\xi(1-\tau)&\geq 0,\nonumber\\
(1-\xi)^2\Lambda+2\xi(1+\tau)&\geq 0,\nonumber\\
-(1-\xi^2)\Lambda+2\xi\tau+2&\geq 0,
\end{align}
where $\tau$ and $\Lambda^{\pm}$ are defined through \Ref{32}.
Now, two general cases occur. Either $0<\xi<1$, or $1<\xi$. First,
take $0<\xi<1$. Then, the inequalities \Ref{31} become
\begin{align}\label{33}
2\tau&\geq -2+(1-\xi)\Lambda^- =:F_1,\nonumber\\
2\tau&\geq -(1+\xi)\Lambda^- =:F_2,\nonumber\\
F_3:=2-(1+\xi)\Lambda^+&\geq 2\tau,\nonumber\\
F_4:=(1-\xi)\Lambda^+&\geq 2\tau,\nonumber\\
2\tau&\geq 2-{{(1+\xi)^2}\over\xi}\Lambda=:F_5,\nonumber\\
2\tau&\geq -2-{{(1-\xi)^2}\over\xi}\Lambda=:F_6,\nonumber\\
2\tau&\geq -{2\over\xi}+{{1-\xi^2}\over\xi}\Lambda=:F_7.
\end{align}
To have solution for $\tau$, $F_3$ and $F_4$ must be greater than
or equal to $F_1$, $F_2$, $F_5$, $F_6$, and $F_7$. So, we have ten
inequalities for $\Lambda^-$ and $\Lambda^+$. The first four,
coming from the $(F_3,\; F_4)\geq(F_1,\; F_2)$, are
\begin{align}\label{34}
4-(1+\xi)\Lambda^+-(1-\xi)\Lambda^-&\geq 0,\nonumber\\
2+(1-\xi)(\Lambda^+ -\Lambda^-)&\geq 0,\nonumber\\
2-(1+\xi)(\Lambda^+ -\Lambda^-)&\geq 0,\nonumber\\
(1-\xi)\Lambda^+ +(1+\xi)\Lambda^-&\geq 0.
\end{align}
The solution to these is the interior of a tetragon  $ABCD$. The
coordinates of its vertices are
$$A\Big(-{{1-\xi}\over{1+\xi}},1\Big),\quad B\Big(1,
{{3+\xi}\over{1+\xi}}\Big),\quad
C\Big({{3-\xi}\over{1-\xi}},1\Big),\quad
D\Big(1,-{{1+\xi}\over{1-\xi}}\Big),$$ where the first coordinate
is $\Lambda^-$, and the second is $\Lambda^+$.

There remains six other inequalities for $\Lambda^-$ and
$\Lambda^+$, to be satisfied. As all of the inequalities are
linear, it is sufficient to check the inequalities on the vertices
of the above tetragon. In fact, we have to compare the values of
$F_3$ and $F_4$, with those of $F_5$, $F_6$, and $F_7$, at the
points $A$, $B$, $C$, and $D$. Doing so, it is seen that the
problem is only with $F_5$ at the point $D$; all other
inequalities are satisfied. At the segments $DA$ and $CD$,
$F_4\leq F_3$. So we have to solve the inequality $F_4\geq F_5$.
The line $F_4=F_5$ passes through $A$, and intersects the segment
$CD$ at the point E:
$$E\Big({{1+6\xi-3\xi^2}\over{(1-\xi)(1+3\xi)}},
-{{1+3\xi^2}\over{(1-\xi)(1+3\xi)}}\Big).$$ So,\\
{\textbf{ii)}}
\begin{equation}\label{35}
0<\xi<1,\quad\mu<0\;(\mu=-1)\quad(\Lambda^-,\Lambda^+)\hbox{
inside the tetragon }ABCE,\quad\tau\hbox{ satisfies \Ref{33}}.
\end{equation}

For the case $1<\xi$, one can still use \Ref{31}. But as $(1-\xi)$
is negative, the first two inequalities in \Ref{33} are reversed.
So we have
\begin{equation}\label{36}
(F_1,\; F_2,\; F_3,\; F_4)\geq 2\tau\geq(F_5,\; F_6,\; F_7).
\end{equation}
From these, one should have
\begin{equation}\label{37}
(F_1,\; F_2,\; F_3,\; F_4)\geq(F_5,\; F_6,\; F_7),
\end{equation}
which are twelve inequalities for $\Lambda^{\pm}$. From $F_1\geq
F_6$ and $F_3\geq F_5$, one obtains
\begin{align}\label{38}
(\xi-1)\Lambda^+ -(\xi+1)\Lambda^- &\geq 0,\nonumber\\
-(\xi-1)\Lambda^+ +(\xi+1)\Lambda^- &\geq 0,\end{align}
respectively. So, one has
\begin{equation}\label{39}
(\xi-1)\Lambda^+ =(\xi+1)\Lambda^- =:\chi.
\end{equation}
This also means that
\begin{equation}\label{40}
F_1=F_3=F_5=F_6=2\tau.
\end{equation}
From $F_1=F_3$, for example, $\chi$ is obtained:
\begin{equation}\label{41}
\chi={{\xi^2-1}\over\xi}.
\end{equation}
It is easily seen that this satisfies \Ref{36}. So, one arrives at\\
{\textbf{iii)}}
\begin{equation}\label{42}
1<\xi,\quad\mu<0\;(\mu=-1)\quad \tau=-{{\xi^2+1}\over{2\xi}},\quad
\Lambda^{\pm}={{\xi\pm 1}\over\xi}.
\end{equation}
It is seen that here there is no allowed region for the rates, but
a single point (apart from scaling) for each value of $\xi$.

Finally, there remains two other special cases. First, the case
$\xi=0$. In this case, the first four inequalities in \Ref{31} are
not changed, and hence the first four inequalities in \Ref{33}.
(One should of course put $\xi=0$ in them.) The vertices of the
tetragon $ABCD$ are now
$$A_0(-1,1),\quad B_0(1,3),\quad C_0(3,1),\quad D_0(1,-1).$$
In the remaining three inequalities of \Ref{31}, there is no
$\tau$, and one reads
\begin{equation}\label{43}
0\leq\Lambda\leq 2.
\end{equation}
It is easy to see that these are satisfied inside the square
$A_0B_0C_0D_0$. This square is in fact the same tetragon $ABCE$ at
the limit $\xi\to 0$. (Note that in this limit, the points $D$ and
$E$ tend to each other.) So,\\
{\textbf{iv)}}
\begin{align}\label{44}
&\xi=0,\quad \mu\ne 0\;\;(\mu=-1),\quad
(\Lambda^-,\Lambda^+)\hbox{ is inside the square
}A_0B_0C_0D_0.\nonumber\\
&\tau\hbox{ satisfies \Ref{31}.}
\end{align}

Finally, in the case $\xi=1$, the first two inequalities in
\Ref{31} become identities, and the sixth and seventh become
identical to each other. So one has four independent inequalities:
\begin{align}\label{45}
-\Lambda^+ +1-\tau\geq 0&,\nonumber\\
-\tau\geq 0&,\nonumber\\
2\Lambda-(1-\tau)\geq 0&,\nonumber\\
1+\tau\geq 0&.
\end{align}
These give the allowed region for $(\Lambda^-,\Lambda^+)$, as the
region limited by the lines $A_1B'$, the horizontal line passing
through $B'$, and the line passing through $A_1$ with the slope
$-1$, containing the point $(1,1)$, where
$$A_1(0,1),\quad B'(0,2).$$
Let us call this region ${\mathbb S}$. So,\\
{\textbf{v)}}
\begin{equation}\label{46}
\xi=1,\quad \mu\ne 0\;\;(\mu=-1),\quad (\Lambda^-,\Lambda^+)\hbox{
is inside }{\mathbb S}.\quad \tau\hbox{ satisfies \Ref{45}.}
\end{equation}
The cases {\textbf i)} to {\textbf v)} summarize the desired
classification.

\vskip\baselineskip

\noindent {\bf Acknowledgement} \\M. Alimohammadi would like to
thank the research council of the University of Tehran, for
partial financial support.

\end{document}